\documentclass[onecolumn,showpacs,amsmath,amssymb,aps,preprint]{revtex4-1}
\usepackage{graphicx}
\usepackage{multirow}
\usepackage{bm}
\usepackage[breaklinks=true,colorlinks=true,linkcolor=blue,urlcolor=blue,citecolor=blue]{hyperref}
\usepackage{booktabs}

\begin{document}
\title{Enhancing the spin transfer torque in magnetic tunnel junctions by ac modulation}

\author{Xiaobin \surname{Chen}$^{1,2}$}
\email{xbchen@physics.mcgill.ca}
\author{Chenyi \surname{Zhou}$^{2}$}
\author{Zhaohui \surname{Zhang}$^{3}$}
\author{Jingzhe \surname{Chen}$^4$}
\author{Xiaohong \surname{Zheng}$^{5,2}$}
\author{Lei \surname{Zhang}$^{6,2}$}
\author{Can-Ming \surname{Hu}$^{3}$}
\author{Hong \surname{Guo}$^{1,2}$}
\email{guo@physics.mcgill.ca}

\affiliation{$^1$ College of Physics and Energy, Shenzhen University, Shenzhen 518060, China\\
             $^2$ Department of Physics, Center for the Physics of Materials, McGill University, Montr′eal, Qu′ebec H3A 2T8, Canada\\
            $^3$ Department of Physics and Astronomy, University of Manitoba, Winnipeg R3T 2N2, Canada\\
            $^4$ Department of Physics, Shanghai University, Shanghai 200444, China\\
            $^5$ Key Laboratory of Materials Physics, Institute of Solid State Physics, Chinese Academy of Sciences, Hefei 230031, China\\
            $^6$ State Key Laboratory of Quantum Optics and Quantum Optics Devices, Institute of Laser Spectroscopy and Collaborative Innovation Center of Extreme Optics, Shanxi University, Taiyuan 030006, China
            }


\begin{abstract}
The phenomenon of spin transfer torque (STT) has attracted a great deal of interests due to its promising prospects in practical spintronic devices. In this paper, we report a theoretical investigation of STT in a noncollinear magnetic tunnel junction under ac modulation based on the nonequilibrium Green's function formalism, and derive a closed-formulation for predicting the time-averaged STT. Using this formulation, the ac STT of a carbon-nanotube-based magnetic tunnel junction is analyzed. Under ac modulation, the low-bias linear (quadratic) dependence of the in-plane (out-of-plane) torque on bias still holds, and the $\sin\theta$ dependence on the noncollinear angle is maintained. By photon-assisted tunneling, the bias-induced components of the in-plane and out-of-plane torques can be enhanced significantly, about 12 and 75 times, respectively. Our analysis reveals the condition for achieving optimized STT enhancement and suggests that ac modulation is a very effective way for electrical manipulation of STT.
\end{abstract}

\maketitle

\section{Introduction.}

When a spin-polarized current flows through a ferromagnetic material, there is a transfer of spin angular momentum near the interface if the spin polarization of the charge carriers is misaligned with that of the ferromagnet. The absorbed components of spin angular momentum of the carriers turn into a torque exerting on the magnetization of the ferromagnet. This is the spin transfer torque (STT) phenomenon that has attracted tremendous interest since its prediction\cite{Slonczewski_JMMM_1996,Berger_PRB_1996} and unambiguous confirmation.\cite{Ralph_Science_1999,Ralph_PRL_2000} Besides extending fundamental insights into spin physics, STT has already been applied in nanoelectronic devices having reduced size and energy consumption.\cite{FertAlbert_NatMat_2007,Brataas_NatMat_2012} Due to STT, a spin-polarized electric current causes precession of magnetization in the ferromagnetic material, and when STT is strong enough, it flips the magnetization direction. Therefore, magnetization can be switched by electric current without the need of any external magnetic field.

Previous studies have revealed many important properties of STT in both metallic and magnetic tunnel junctions (MTJs). STT can be in-plane and out-of-plane. Usually, the in-plane STT is proportional to ${\bf{\hat m}} \times ( {{\bf{\hat M}} \times {\bf{\hat m}}} )$ \cite{Slonczewski_JMMM_1996} where the vectors ${{\bf{\hat M}}}$ and ${\bf{\hat m}}$ are magnetizations of the fixed- and free-ferromagnets in the MTJ, respectively. By comparison, the field-like or out-of-plane STT is proportional to ${{\bf{\hat m}} \times {\bf{\hat M}}}$, which is attributed to interlayer exchange coupling intermediated by tunneling electrons between the two noncollinear ferromagnets\cite{Slonczewski_PRB_1989,ASchuhl_PRB_2002}. Generally, the out-of-plane torque is negligible in metallic junctions.\cite{Ralph_JMMM_2008} For practical applications, it is important to consider the bias dependence of STT. The in-plane/out-of-plane torque has a good linear/quadratic dependence on bias when bias is small, as revealed by its derivative relative to the applied bias from ferromagnetic resonance experiments.\cite{Slonczewski_Nat_2007} The bias dependence also varies with structural parameters.\cite{Theodonis_PRL_2006,Datta_IEEE_2012} STT is also found to be affected by other factors, including layer index,\cite{Guo_PRB_2007,XiaKe_PRB_2008} disorder scattering\cite{XiaKe_PRB_2008}, asymmetry electrodes\cite{Datta_IEEE_2012}, and so on.

Achieving high-efficiency STT devices is very important for application, and this turns out to be a global challenge. Operating at elevated bias could increase STT, but high bias is usually undesirable. Theoretically, this shows that when only ac bias is present, the in-plane STT sharply increases in MTJs where the ferromagnetic leads are separated by a vacuum.\cite{SuGang_PRB2003} Certain interfacial disorder could slightly increase STT in ferromagnetic spin valves\cite{XiaKe_PRB_2008} while magnifying STT in Fe/MgO/Fe MTJs\cite{XiaKe_PRB_2011}. Hatami $et~al.$ predicted that thermal generation could lead to huge out-of-plane STT in metallic spin valves, although such a phenomenon is still under experimental exploration.\cite{Hatami_PRL_2007}. Recently, it was revealed that spin-orbit coupling may act as another mechanism to efficiently manipulate current-induced torques.\cite{Chernyshov_NatPhys_2009,Brataas_NatMat_2012,FanYabin_NatMater_2014} An interesting possibility that has not been investigated so far, is if STT can be enhanced by applying an external ac modulation (without increasing the total bias) to the MTJ - although it has been known that such ac modulation can increase charge current flow\cite{cxb_prb2013_ac}.

It is the purpose of this work to report theoretical investigations of STT manipulation by ac harmonic modulation. Experimentally, such modulation can be achieved by applying ac modulation signals or light irradiation. With ac modulation turned on or off, we found that the system can be switched to a ``write" or ``read" state, thus ac modulation offers an elegant and efficient control to STT-MTJ devices. Based on the Keldysh nonequilibrium Green's function (NEGF) formalism, we formulate and derive in closed form the time-averaged STT under ac modulation. To illustrate the idea, we further analyze carbon nanotube (CNT) MTJs under ac modulation: such CNT MTJ can be well described by a tight-binding atomic model\cite{Guo_PRL_2000}, and it was also realized experimentally\cite{Aurich_APL_2010}. Our calculation indicates that opening more transport channels by ac modulation at designated frequency $\omega$ enhances STT: both in-plane and out-of-plane bias-induced STTs exhibit significant enhancement by up to 12 and 75 times, respectively. Analytically we predict that the STT enhancement achieves peak values when the ac modulation amplitude $\Delta$ and frequency $\omega$ are set such that $\Delta/\omega$ is around extreme points of a Bessel function (see below). Our theory reveals an exciting mechanism that STT can be controllably engineered via ac modulation.

The rest of the paper is organized as follows. In Sec.~\ref{sec:theory} we derive the formulation of the ac modulated STT. Sec.~\ref{sec:results&discussion} represents numerical results of a CNT-based MTJ and related discussions. Finally Sec.~\ref{sec:conclusion} presents a conclusion of this work.

\section{Theory\label{sec:theory}}
    \subsection{Time-averaged ac spin transfer torque}
    \begin{figure*}
            \includegraphics[width=0.9\textwidth]{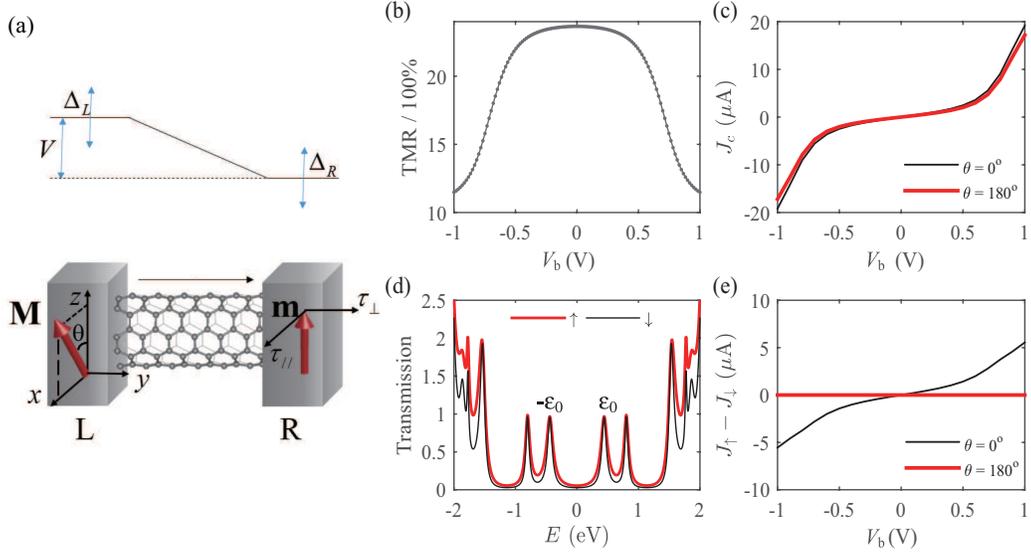}
            \caption{Schematic plot of a CNT-based MTJ device and its dc magneto-resistive behaviors. (a) In the upper panel, a static bias accompanied by harmonic modulations is also shown. In the lower panel, a carbon nanotube is sandwiched by two metallic electrodes, which are marked as the lead L and R, respectively. The unit magnetization vector of the lead L, $\bf\hat{M}$, is fixed and lies within the $xz$ plane, while that for lead $R$, $\bf \hat{m}$, orients along the $z$ axis to facilitate further analysis for spin transfer torques. The black arrow placed above the carbon nanotube shows the direction of positive charge current.
            For an MTJ based on a (5,5) CNT of 5 unit-cell length with $g_{\uparrow}=0.5$ eV and $g_{\downarrow}=0.25$ eV [see Eqs. (\ref{eq:GammaR}) and (\ref{eq:GammaL})], its (b) TMR as a function of dc bias $V_b$, (c) charge current under parallel ($\theta=0^\circ$) and anti-parallel ($\theta=180^\circ$) configurations, (d) zero-bias transmission spectra of spin-up and spin-down electrons at $\theta=0^\circ$, and (e) spin-$z$ current as a function of bias.          }\label{fig1}
    \end{figure*}


        As illustrated in Fig.~\ref{fig1} (a), we consider a magnetic tunnel junction with noncollinear ferromagnetic metallic leads, which are assumed to be reservoirs with chemical potentials $\mu_\textrm{L}$ and $\mu_\textrm{R}$. The system is under a dc bias $V_{b}$ with a time-dependent harmonic modulation of amplitude $\Delta_{\textrm{L,R}}$. For simplicity of further analysis, we suppose that the transport direction of the system is along the $y$ axis, and that magnetization of the left lead (with fixed magnetization) and right lead (with free magnetization) point within the $xz$ plane and along the $z$ direction, respectively.
        The Hamiltonian of the system can be written as $\hat {H}=\hat{H}_\textrm{L}+\hat{H}_\textrm{R}+\hat{H}_\textrm{C}+\hat{V}$, with\cite{cxb_prb2013_ac,ChenJingzhe_jpcm_2014,YuYunjin_Nanotech2013}
        \begin{align}
            {\hat{H}_L} &= \sum\limits_{k,s=\pm }   \left\{ \left[ \varepsilon _{k L s  }(t)+qV_{b}/2+ s  {M_L}\cos \theta  \right]   \hat{c}_{k L s  }^\dag {\hat{c}_{k L s  }} \right. \cr
                   & \left. + {M_L}\sin \theta \hat{c}_{k L s  }^\dag \hat{c}_{k L \bar s }\right\},  \\
             {\hat{H}_\textrm{R}} &= \sum\limits_{ks  } {\left[ {{\varepsilon _{k \textrm{R} s  }(t)} -qV_{b}/2+ s  {M_\textrm{R}}} \right] \hat{c}_{k \textrm{R} s  }^\dag { \hat{c}_{k \textrm{R} s  }}} ,\\
            {\hat{H}_\textrm{C}} &= \sum\limits_{m,s  } {[{\varepsilon _m}+qV(y_m)] \hat{d}_{ms  }^\dag { \hat{d}_{ms  }}}  + \sum\limits_{ <m,n>, s  } {\gamma \hat{d}_{ms  }^\dag {\hat{d}_{ns  }}} ,\label{eq:Hc}\\
            \hat{V} &= \sum\limits_{ s  ,n; k\alpha  \in \textrm{L,R}} {{t_{k\alpha ,n}} \hat{c}_{k\alpha s  }^\dag {\hat{d}_{ns  }} + \textrm{H.c.}} ,
        \end{align}
        where ${\varepsilon _{k\alpha s }}(t) = {\varepsilon _{k\alpha s }^0} + {\Delta _\alpha }\cos \omega t ~ (\alpha \in L,R)$ represents the effect of ac harmonic modulation on leads, $q=-e$ is the electron charge, $M_{\textrm{L(R)}}$ is the total magnetic moment of lead L(R) with $\bf{\hat M} (\bf{\hat m})$ being the unit magnetization vector, $V_b$ is dc bias voltage, and $V(y)= V_{b}/2 - (V_{b}/L_c)y$ with $y=0$ being the position of the left contact surface and $L_c$ being the length of the sandwiched scattering region.
        Also, ${ \hat{c}_{k\alpha s }}( \hat{c}_{k\alpha s }^\dag)$ annihilates (creates) an electron in lead $\alpha$ labeled by $k$ and spin $s$ ($s  =  + ,- $), and ${\hat{d}_{ns }}(\hat{d}_{ns }^\dag)$ annihilates (creates) an electron with spin $s$ at site $n$ in the central region. $\gamma$ describes the nearest-neighbor hopping integral in the central region, $<m,n>$ means that $m$ and $n$ are nearest-neighbor sites, and ${t_{k\alpha ,n}}$ represents interaction between leads and the central region.

        In this MTJ device, the lead L acts as a spin polarizer, which injects spin-polarized current into the central region. When going through the lead $R$, the spin polarization direction of the carriers generally aligns with the magnetization direction of lead $R$, indicating spin relaxation and a corresponding loss of spin angular momentum. Spin relaxation in ferromagnetic materials is really fast; for example, the characteristic length in transition metals is less than 1 nm.\cite{FertAlbert_NatMat_2007,Guo_PRB_2007,Flipse_PRB_2014} Due to conservation of spin angular momentum, the loss in spin currents leads to an effective torque acting on $\hat{\bf  m}$. Therefore, the so-called spin transfer torque is intrinsically an interfacial effect and can be calculated using the spin currents perpendicular to $\hat{\bf  m}$ to a good approximation.\cite{Stiles_PRB_2002Anatomy} As shown in Fig.~1 (a), the spin transfer torques acting on $\bf  m$ can be decomposed into two components: out-of-plane STT, $\tau_{\perp}$, and in-plane STT, $\tau_{||}$, and they can be calculated as the $x$ and $y$ components of the spin current flowing into lead $R$ as:\cite{Theodonis_PRL_2006}
        \begin{align}
        {\tau _{||}} &= J_x^s,  \label{taull}\\
        {\tau _ \bot } &= J_y^s,\label{taun}
        \end{align}
        respectively.

        To get spin currents from the central region to the lead $R$, we can calculate the hopping part from the time evolution of the spin operator in the lead $R$,\cite{Theodonis_PRL_2006,Guo_PRB_2007,Ralph_JMMM_2008}
        \begin{align}\label{eq:def}
           {\bf{J}}_{\textrm{C} \to \textrm{R}}^s\left( t \right) ={\left\langle {\frac{{d{\hat{{\bf{S}}}_\textrm{R}}}}{{dt}}} \right\rangle _{hopping}},
        \end{align}
        where $\hat{{\bf{S}}}_\textrm{R} = \sum\limits_{i \in \textrm{R}}^{} {{\hat{{\bf{s}}}_i}}$ and ${{\bf{\hat {s}}}_i} = \frac{\hbar }{2}\sum_{ss '} {\hat{ c}_{is }^\dag {{\bm{ \hat{\sigma} }}_{ss'}}{\hat{ c}_{is'}}} $ with Pauli matrix ${\bm{\hat {\sigma} }} = \left( {{\sigma _x},{\sigma _y},{\sigma _z}} \right)$.\cite{Ralph_JMMM_2008,XingDY_PRL_2010} This definition of spin angular momentum is actually equivalent to $\hat{s}_i^u  = \hat{ C}_{i + }^\dag {\hat{ C}_{i +  }} - \hat{ C}_{i - }^\dag {\hat{ C}_{i -}}$, where $u=x,y,z$ and $\hat{ C}_{i +(-) }(\hat{ C}^\dag_{i +(-)} )$ is annihilation (creation) operator of the spin eigenstates of the local spin quantization axis along the $u$ direction.
        \textcolor{blue}{The equation of motion for spin angular momentum consists of two parts: spin current flux contributed by hopping and precessional time evolution of spins under the influence of effective on-site magnetic fields.\cite{Guo_PRB_2007} Here, the $\left\langle {d{\hat{{\bf{S}}}_\textrm{R}}}/{dt} \right\rangle _{hopping}$ in Eq.~(\ref{eq:def}) means that we keep only the hopping contribution.}

        Spin currents flowing into the lead $R$ from the central region are (From here on, we set $e=1,\hbar =1 $)(See Appendix \ref{app1} for a detailed derivation)
        \begin{align}
            {\bf{J}}_{\textrm{C} \to \textrm{R}}^s  (t)=  - \sum\limits_{ s  s  ',k\alpha  \in \textrm{R}, n\in \textrm{C}} {{\rm{Re}}\left[ {G_{ns  ,k\alpha s  '}^ < \left( {t,t} \right){t_{k\alpha,n}}{{\bm{\sigma  }}_{s's  }}} \right] }.
        \end{align}

        Using Dyson's equation and analytic continuation rules\cite{Jauho_PRB_1994} for the Green's function of lead $R$, we have
        \begin{align}
            G_{ns ,k\alpha s '}^ < \left( {t,t} \right)
             &= \sum\limits_{n'} \int {d{\tau _1}} G_{ns ,n's '}^r\left( {t,{\tau _1}} \right)t_{n',k\alpha }g_{k\alpha s '}^ < \left( {{\tau _1},t} \right) \cr
            &+ \sum\limits_{n'}\int {d{\tau _1}} G_{ns ,n's '}^ < \left( {t,{\tau _1}} \right){t_{n',k\alpha }}g_{k\alpha s '}^a\left( {{\tau _1},t} \right),
        \end{align}
        which combined with the Green's functions for isolated lead $R$,\cite{Jauho_PRB_1994}
        \begin{align}
            g_{k\alpha s }^ < \left( {{\tau _1},t} \right)
            &= if\left( {\varepsilon _{k\alpha s }^0} \right){e^{ - i\varepsilon _{k\alpha s }^0\left( {{\tau _1} - t} \right)}}{e^{ - i\int_t^{{\tau _1}} {{\Delta _R }\left( \tau  \right)d\tau } }}, \\
            g_{k\alpha s }^a\left( {{\tau _1},t} \right)
            &= i\theta \left( {t - {\tau _1}} \right){e^{ - i\varepsilon _{k\alpha s }^0\left( {{\tau _1} - t} \right)}}{e^{ - i\int_t^{{\tau _1}} {{\Delta _R }\left( \tau  \right)d\tau } }},
        \end{align}
        leads to
        \begin{align}
            {\bf{J}}_{\textrm{C} \to \textrm{R}}^s\left( t \right)
            & =   {\mathop{\rm Im}\nolimits} \sum\limits_{s s ',nn'} \int_{} {\frac{{d\varepsilon }}{{2\pi }}} \int_{ - \infty }^t {d{\tau _1}} {e^{ - i\varepsilon \left( {{\tau _1} - t} \right)}}{e^{ - i\int_t^{{\tau _1}} {{\Delta _\textrm{R}}\left( \tau  \right)d\tau } }} \cr
            & \cdot \left[ {G_{ns ,n's '}^r\left( {t,{\tau _1}} \right){f_\textrm{R}}\left( \varepsilon  \right)} \right. \cr
            & + \left. {G_{ns ,n's '}^ < \left( {t,{\tau _1}} \right)} \right] {\Gamma _{\textrm{R};n's ',ns '}} \left( \varepsilon  \right) {{\bm{\sigma }}_{s 's }}.
        \end{align}
        Here the replacement ${\sum\nolimits_{k\alpha} { \to \int_{} {d\varepsilon  \cdot {\rho _{\alpha s} }\left( \varepsilon  \right)} } }$ , where $ {\rho _{\alpha s} }\left( \varepsilon  \right)$ is spin-resolved density of states of the lead $\alpha$, is used, and the static bandwidth function of lead $R$ is defined as
        \begin{align}
            {{\Gamma _{\textrm{R};n's ',ns '}}\left( \varepsilon  \right) \equiv 2\pi \sum_{\alpha \in \textrm{R}} \rho_{\alpha,s'} \left( \varepsilon  \right)t_{n',\alpha}\left( \varepsilon  \right){t_{\alpha,n}}\left( \varepsilon  \right)}.
        \end{align}

        To get a simpler expression, we use the wide-band limit, where the real parts of self-energies are neglected and the energy dependence of the imaginary parts are presumed to be weak enough to be ignored.\cite{cxb_prb2013_ac} Under this assumption, we have
        \begin{widetext}
        \begin{align}\label{eq:Jc}
            {\bf{J}}_{\textrm{C} \to \textrm{R}}^s\left( t \right)
            &=   {\mathop{\rm Im}\nolimits} \sum\limits_{s s ',nn'} \left[ {\int_{} {\frac{{d\varepsilon }}{{2\pi }}} \int_{ - \infty }^t {d{\tau _1}} {e^{ - i\varepsilon \left( {{\tau _1} - t} \right)}}{e^{ - i\int_t^{{\tau _1}} {{\Delta _R }\left( \tau  \right)d\tau } }}G_{ns ,n's '}^r\left( {t,{\tau _1}} \right){f_\textrm{R}}\left( \varepsilon  \right)} \right. \cr
            &+ \frac{1}{2}\left. {G_{ns ,n's '}^ < \left( {t,t} \right)} \right]{\Gamma _{\textrm{R};n's ',ns '}} {{\bm{\sigma }}_{s ',s }}\cr
            &={\mathop{\rm ImTr}\nolimits} \left\{ \left[ {\int_{}^{} {\frac{{d\varepsilon }}{{2\pi }}} {A_\textrm{R}}\left( {\varepsilon ,t} \right){f_\textrm{R}}\left( \varepsilon  \right)}
             + \frac{1}{2} {\sum\limits_{\alpha  \in \textrm{L,R}} i \int {\frac{{d\varepsilon }}{{2\pi }}} {f_\alpha }\left( \varepsilon  \right){A_\alpha }\left( {\varepsilon ,t} \right){{\Gamma }_\alpha }A_\alpha ^\dag \left( {\varepsilon ,t} \right)} \right]{\Gamma _\textrm{R}}{\bm{\sigma }} \right\}
        \end{align}
        \end{widetext}

        with
        \begin{align}
            {A_\alpha }\left( {\varepsilon ,t} \right) \equiv \int_{ - \infty }^{ + \infty } {d{\tau _1}} {G^r}\left( {t,{\tau _1}} \right){e^{ - i\varepsilon \left( {{\tau _1} - t} \right)}}{e^{ - i\int_t^{{\tau _1}} {{\Delta _\alpha }\left( \tau  \right)d\tau } }}.
        \end{align}

        Further under time average, \cite{cxb_prb2013_ac}

        \begin{align}
        &\left\langle {{A_\alpha }\left( {\varepsilon ,t} \right)} \right\rangle
        = \sum\limits_k {J_k^2\left( {\frac{{{\Delta _\alpha }}}{\omega }} \right)} {G^r}\left( {\varepsilon  - k\omega } \right),\label{eq:Aave}\\
        &\left\langle {{A_\alpha }\left( {\varepsilon ,t} \right){{{\Gamma }}_\alpha }A_\alpha ^\dag \left( {\varepsilon ,t} \right)} \right\rangle \cr
        &= \sum\limits_k {J_k^2\left( {\frac{{{\Delta _\alpha }}}{\omega }} \right)} {G^r}\left( {\varepsilon  - k\omega } \right)  {{{\Gamma }}_\alpha }{G^a}\left( {\varepsilon  - k\omega } \right),\label{eq:AGAave}
        \end{align}
        where $J_k$ is $k^\textrm{th}$-order Bessel function of the first kind, $\Gamma=\Gamma_L+\Gamma_R$, and the retarded/advanced Green's function $G^{r/a}(\varepsilon)$ is defined as\cite{cxb_prb2013_ac,ZhouChenyi_PRB2016}:
        \begin{align}
        {G^{r/a}}\left( \varepsilon \right) = {\left( {\varepsilon \pm i{0^ + } - {H_C} \pm \frac{i}{2}\Gamma } \right)^{ - 1}},
        \end{align}
        These two Green's functions are actually the retarded and advanced Green's functions of the steady-state system, i.e., without ac modulation. The coincidence inherited from the wide-band limit greatly simplifies our problem by representing quantities using steady-state Green's functions. Thus,
        taking time average of Eq.~(\ref{eq:Jc}) and using Eqs.~(\ref{eq:Aave})-(\ref{eq:AGAave}), we have the time-averaged quasi-ballistic spin current as:
        \begin{align}\label{eq:aveJs}
        &\left\langle {J_{\textrm{C} \to \textrm{R};\nu }^s} \right\rangle
         = \int_{}^{} {\frac{{d\varepsilon }}{{2\pi }}} \sum\limits_k {J_{k;\textrm{R}}^2} {\rm{ImTr}}\left( {G_k^r{\Gamma _\textrm{R}}{{\bf{\sigma }}_{\bf{\nu }}}} \right){f_\textrm{R}}\left( \varepsilon  \right)\cr
        & + \sum\limits_{\alpha  \in \textrm{L,R}} {} \int {\frac{{d\varepsilon }}{{4\pi }}} {f_\alpha }\left( \varepsilon  \right)\sum\limits_k {J_{k;\alpha}^2} {\rm{ReTr}}\left( {G_k^r{\Gamma _\alpha }G_k^a{\Gamma _\textrm{R}}{\sigma _\nu }} \right),
        \end{align}
        with
        \begin{align}
        G_{  k}^r \equiv {G^r}\left( {\varepsilon  - k\omega } \right),J_{k;\alpha} \equiv J_{k;\alpha}\left( {\Delta_\alpha /\omega } \right) .
        \end{align}

        Further simplification under our coordination leads to(See Appendix \ref{app2})
        \begin{align}
        J_{x/z}^s
        & = \int_{}^{} {\frac{{d\varepsilon }}{{4\pi }}} \sum\limits_k {\left( {{f_L}J_{k;L}^2 - {f_\textrm{R}}J_{k;\textrm{R}}^2} \right)} {\rm{Tr}}\left( {G_k^r{\Gamma _L}G_k^a{\Gamma _\textrm{R}}{\sigma _{x/z}}} \right),
        \end{align}
        which shows a rectification effect as seen in microwave experiments\cite{Bauer_PRB_2015} that spin currents are nonzero under pure ac bias, i.e., $V_b=0$ and $\Delta_L\ne\Delta_\textrm{R}$.
        When $\Delta_L=\Delta_R$, specifically, there is no ac bias and the total bias applied in the system does not change. In this case, the time-averaged in-plane spin currents can be simplified to a Caroli-like formula:
        \begin{align}\label{eq:Js_xzSim}
              J_{x/z} ^s =\int_{}^{} {\frac{{d\varepsilon }}{{4\pi }}} \left( {{f_L} - {f_\textrm{R}}} \right)
              \sum\limits_k {J_k^2} \textrm{Tr}\left( {G_{  k}^r{\Gamma_L}G_{  k}^a{\Gamma_\textrm{R}}{{\rm{\sigma }}_{x/z} }} \right),
        \end{align}
        while the time-averaged out-of-plane spin current is
        \begin{align}\label{eq:Js_ySim}
                 J_y ^s =   \int_{}^{} {\frac{{d\varepsilon }}{{2\pi }}} \sum\limits_k {J_{k}^2} {\rm{ImTr}}\left( {G_{ k}^r {\Gamma _\textrm{R}} {{\bf{\sigma }}_y}} \right){f_\textrm{R}}\left( \varepsilon  \right).
        \end{align}
         Without causing confusion, here we omit the average brackets ``$\langle\rangle$". Note that this simplification requires a wide-band limit, the coordinates defined in Fig.~\ref{fig1}, and particularly that the magnetization of lead $R$ aligns along the $z$ axis. According to Eqs.~(\ref{taull}) and (\ref{taun}), we directly obtain the time-averaged in-plane and out-of-plane STTs. It is worth noting that Eq.~(\ref{eq:Js_ySim}) shows that the out-of-plane torque is contributed by all valence bands and should be nonzero even under zero bias, where there is no electric current flowing in the system. This static term represents an effective precession associated with interface coupling, contributed by all occupied states. Exact evaluation of this term requires information of all bands that are around and below the Fermi energy. However, the static term doesn't manifest in ferromagnetic resonance detections of STTs.\cite{Slonczewski_Nat_2007} Therefore, it is reasonable to consider the bias-induced component of the out-of-plane STT only. The bias-induced portion of the out-of-plane STT mainly comes from the transport channels around the bias window, and thus is well-defined. In the following, we shall focus on bias-induced components. The calculated out-of-plane value subtracts the corresponding zero-bias value, i.e., $\tau(V)-\tau(0)$. Without causing ambiguity, bias-induced STTs are referred to simply as STTs.

    \subsection{CNT-based MTJs}
        Using these formulas, we further investigate the ac modulation of STTs using a carbon nanotube (CNT) as the scattering region. This provides a simple and clear demonstration of our proposal of ac modulation as a tuning knot for spin transfer torques in MTJs. As usual, CNTs can be described using the nearest-neighbor tight-binding method with the hopping integral $\gamma=-2.6~$eV,\cite{cxb_valley_2015,cxb_PRB_2014} shown as the second term in Eq.~(\ref{eq:Hc}). The bandwidth function $\Gamma_{\textrm{R}}$ can be obtained as\cite{Guo_PRL_2000}:
        \begin{align}\label{eq:GammaR}
        \Gamma _{\textrm{R}; mn} =\left\{ \begin{array}{ll}
        {\delta _{mn}}\left( {\begin{array}{*{20}{c}}
        {{ g_{\uparrow}}}&{}\\
        {}&{g_{\downarrow} }
        \end{array}} \right),& {\rm{ if~site~}} n {\rm{~is~adjacent}}\\
        & {\rm{to~lead~R,}}\\
        0,& \rm{otherwise.}
        \end{array} \right.
        \end{align}
        $g_{\uparrow/\downarrow}$ reflects the interaction strength of spin-up/spin-down electrons between the lead $R$ and the central region, while the bandwidth function of lead $L$, whose magnetization has an angle of $\theta$ to the axis $z$, is
        \begin{align}\label{eq:GammaL}
        \Gamma _{L} = {{\mathcal{R}}^\dag }\Gamma _{L}^{0}\mathcal{R},
        \end{align}
        with the rotation matrix
        \begin{align}
        \mathcal{R} = \left( {\begin{array}{*{20}{c}}
        {\cos (\theta/2) }&{\sin (\theta/2) }\\
        { - \sin (\theta/2) }&{\cos (\theta/2) }
        \end{array}} \right),
        \end{align}
        and $\Gamma _{L}^{0}$ in the same form with $\Gamma _{\textrm{R}}$,
        \begin{align}\label{eq:GammaL0}
        \Gamma _{L; mn}^0 =\left\{ \begin{array}{ll}
        {\delta _{mn}}\left( {\begin{array}{*{20}{c}}
        {{ g_{\uparrow}}}&{}\\
        {}&{g_{\downarrow} }
        \end{array}} \right), & {\rm{ if~site~}} n {\rm{~is~adjacent}}\\
        &  {\rm{to~lead~L,}}\\
        0,&\rm{otherwise.}
        \end{array} \right.
        \end{align}
        Here we simply assume that lead $L$ and lead $R$ are identical by using the same $g_{\uparrow}$ and $g_{\downarrow}$ in both leads.

    \begin{figure}
        \includegraphics[width=0.48\textwidth]{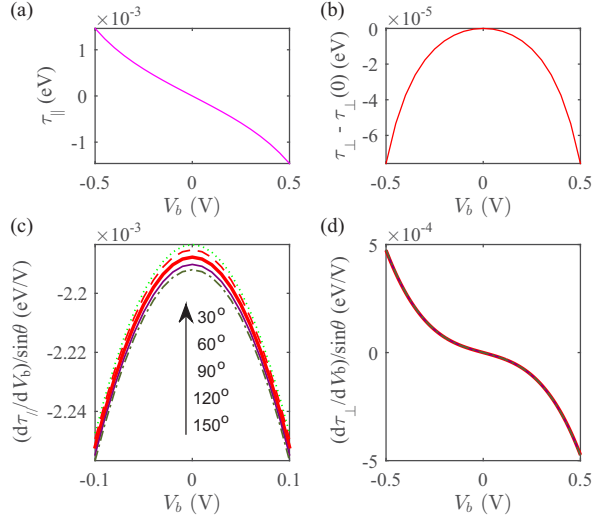}
        \caption{Dc STT properties of a (5,5)CNT-N5 MTJ. Bias dependence of the (a) in-plane and (b) out-of-plane STTs when $\theta=90^\circ$, and (c) in-plane and (d) out-of-plane torkances scaled by $\sin\theta$ under different noncolinear angles: $\theta$=150$^\texttt{o}$ (olive dash-dotted line), 120$^\texttt{o}$ (purple line), 90$^\texttt{o}$ (thick red line), 60$^\texttt{o}$ (thin red dashed line), 30$^\texttt{o}$ (green dotted line). }
        \label{fig2}
    \end{figure}

\section{Results and Discussion\label{sec:results&discussion}}

    \begin{figure}
        \includegraphics[width=0.48\textwidth]{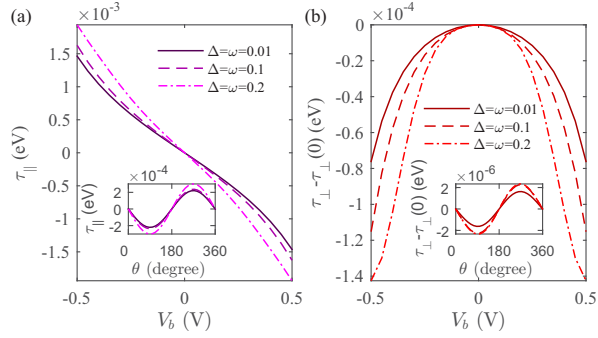}
        \caption{STTs under ac modulation of $\Delta_L=\Delta_\textrm{R}\equiv \Delta$ in the FM/(5,5)CNT-N5/FM MTJ device. (a) In-plane and (b) out-of-plane STTs as functions of dc bias $V_b$ with $\theta=90^\circ$, and of noncollinear angle $\theta$ with $V_b=0.01$ V (insets) when $\Delta=\omega=0.01$ (solid line), 0.1(dash line), 0.2(dash dotted line) eV. }\label{fig3}
    \end{figure}

    To show the dc magnetoresistive performance of CNT-based MTJs, we present the results of dc transport calculation of an MTJ using a 5-unit-cell CNT with index (5,5) as the scattering region, labeled as (5,5)CNT-N5. $g_\uparrow=0.5$~eV and $g_\downarrow=0.25$ eV in both leads, which correspond to about 33\% spin polarization in leads (such as Co\cite{Tedrow_PRB_1973}). Although an ideal periodic (5,5) CNT is metallic, a finite sample possesses discrete energy levels and the (5,5)CNT-N5 structure has an energy gap of about 0.9 eV; further
     sandwiching a finite sample between two metallic leads can make it a good tunnel magnetoresistance (TMR) device.\cite{Guo_PRL_2000} As shown in Fig.~\ref{fig1} (b), the zero-bias TMR of the device reaches about 23\%. Here, TMR is defined as $\textrm{TMR}=(J_{\textrm{P}}-J_{\textrm{AP}})/J_{\textrm{AP}}$, using currents under parallel (P) and anti-parallel (AP) configurations. As bias increases, TMR decreases. However, TMR of this device remains above 10\% even when the dc bias voltage ($V_b$) reaches 1 V, and the fractional reduction at a bias of 0.5 V is only 2\%, much better than the 72\% reduction of Fe/MgO/Fe MTJ devices.\cite{Slonczewski_Nat_2007}

     Fig.~\ref{fig1} (c) shows the response of charge current $J_c$ to bias voltage $V_b$ under P and AP configurations. As expected, charge current $J_c$ has a higher output under a P configuration than under an AP configuration. They both have a semiconducting behavior, where $J_c$ increases linearly as bias increases under a small bias, and they have a significant turning point around $V_b=\pm$0.5 V. The turning points indicate the involvement of new transport channels contributed by resonant tunneling. This is evidenced by the spin-resolved transport spectrum under a P configuration and zero bias in Fig.~\ref{fig1} (d), where both spin-up and spin-down electrons exhibit transmission peaks at around $\pm$0.44 eV (marked as $\pm\varepsilon_0$ in the plot) away from the Fermi energy.
    Due to this feature, the bias dependence of the P-configuration spin-$z$ current resembles that of charge current, as shown in Fig.~\ref{fig1} (e). Under an AP configuration, the charge current of spin-up and spin-down electrons is the same, resulting in zero net spin-$z$ current.
    Also, the output charge current is in the order of $\mu$A. For a given area of around 46 \AA$^2$ [the diameter of a (5,5) CNT is about 6.8 \AA], the current density is about $1\times10^8~$A/cm$^2$. To meet the requirement of switching current densities in permalloy\cite{Bauer_PRB_2010} and MgO-based MTJs devices, which range from $1\times 10^6$ to $1\times 10^9$ A/cm$^2$,\cite{Ralph_Science_1999,DiaoZhitao_JPCM_2007,Ralph_JMMM_2008,Kubota_NatPhys_2008} the area of the ferromagnetic leads can be at most a dozen times larger than that of the CNT's, which shows a promising potential for this kind of MTJ device.

    As stated in Sec. II, when the magnetization of leads $L$ and $R$ is collinear, the loss of spin angular momenta during injection of electrons from one lead to another is also collinear, and thus, there will be no spin transfer torque. In other words, spin transfer torques originate from spin currents that are perpendicular to the direction of magnetic moments in the free magnet, $\hat{ \bf m}$. In our setup, the corresponding in-plane and out-of-plane STTs are contributed from spin-$x$ and spin-$y$ currents, respectively.

    \begin{figure}
        \includegraphics[width=0.45\textwidth]{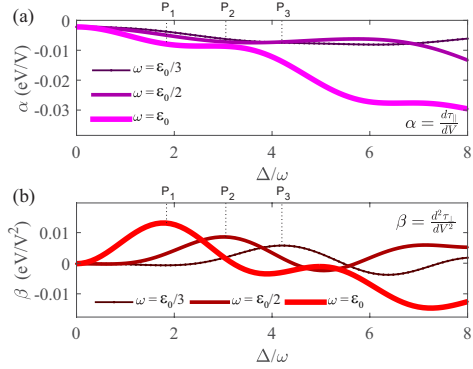}
        \caption{Ac modulation on a FM/(5,5)CNT-N5/FM device with $\theta=90^\circ$ under ac modulation of $\Delta_L=\Delta_R\equiv \Delta$. Low-bias coefficients of the (a) in-plane and (b) out-of-plane STTs as a function of $\Delta/\omega$, with $\omega=\varepsilon_0/3$ (thin lines),$\varepsilon_0/2$(thick lines), and $\varepsilon_0$(very thick lines), respectively. Positions of the first peak of $J_k(x)$ ($k=1,2,3$), $P_k$, are also shown. }\label{fig4}
    \end{figure}

    \begin{figure*}
        \includegraphics[width=0.8\textwidth]{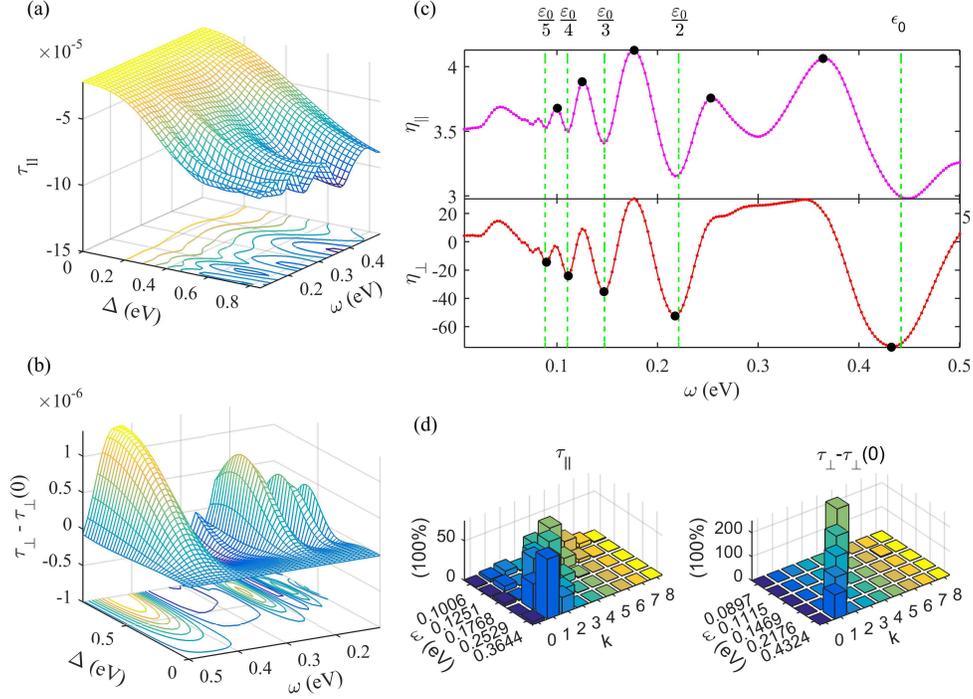}
        \caption{Ac modulation on a FM/(5,5)CNT-N5/FM device with $\theta=90^\circ$ under ac modulation of $\Delta_L=\Delta_R=\Delta$ and dc bias $V_b=0.01$~V.
        Contour plot of the (a) in-plane and (b) out-of-plane STTs as functions of $\omega$ and $\Delta$.
         (c) Enhancement factors of the in-plane (upper panel) and out-of-plane (lower panel) STTs as a function of $\omega$ under $\Delta=0.619$ eV. The positions of $\varepsilon_0/k,~k=1,...,5$ are also shown. (d) Decomposed contributions from $k$-photon-assisted tunneling at those marked points in (c). The unit of STTs shown in the figure is eV.
         }\label{fig5}
    \end{figure*}
    To gain a clear picture about STTs in this CNT-based device, we present its dc properties of STTs in Figs.~\ref{fig2}(a) and (b).
    Under a small bias within 0.1 V, the in-plane and out-of-plane STTs demonstrate a common linear and quadratic dependences on bias, respectively, as predicted by Slonczeweski\cite{Slonczewski_JMMM_1996} and as those detected in Fe/MgO/Fe MTJs.\cite{Slonczewski_Nat_2007,Kubota_NatPhys_2008}
    Theodonis $et~al.$ pointed out that an anomalous bias dependence of the in-plane STT arises by tuning energy levels for the ferromagnetic materials.\cite{Theodonis_PRL_2006} Wilczy\'nski $et~al.$ further showed that the bias dependence of the in-plane STT is asymmetric even for symmetry junctions.\cite{Wilczynski_PRB_2008} And Datta $et~al.$ explained the voltage asymmetry observed in experiments by energy dependence in the spin polarization of leads.\cite{Datta_IEEE_2012} Here, the in-plane and out-of-plane STTs show perfect symmetry and anti-symmetry about the bias, respectively, which can be attributed to the electron-hole symmetry of our device. It is also shown in the figure that a complicated bias dependence emerges at larger bias,\cite{Slonczewski_Nat_2007,XiaKe_PRB_2011,Kubota_NatPhys_2008} where the in-plane torque shows a substantial increase around 0.1 V.

    In addition, the bias-induced in-plane component is two orders larger than the out-of-plane one. In metallic systems,
    the magnitude of the out-of-plane component is 1-3\% of the magnitude of the in-plane component.\cite{Ralph_JMMM_2008}
    Meanwhile, it is shown to be comparable to the in-plane one as evidenced in experiments\cite{Slonczewski_Nat_2007,Kubota_NatPhys_2008,Brataas_NatMat_2012} and from
    theoretical investigations in MgO-based MTJs.\cite{Theodonis_PRL_2006,Wilczynski_PRB_2008,Heiliger_PRL_2008}
    Here, the out-of-plane STT may be under-estimated because of two reasons. One reason is the usage of a wide-band limit, where the second term in Eq.~(\ref{eq:aveJs}) is omitted. The other reason is the single-orbital tight-binding model, where only the portion from $\pi$ orbitals is counted. Nevertheless, as we shall see in the following the out-of-plane STT shows the basic features of photon-assisted tunneling.

    Besides the distinct bias dependence, both torques are expected to have the same angular dependence as proportional to $\sin \theta$,\cite{Slonczewski_JMMM_1996,Alexey_PRB_2007,Heiliger_PRL_2008} In Figs.~\ref{fig2}(c) and (d), we draw in-plane and out-of-plane torkances, i.e., $d\tau_{||,\perp}/dV_b$, scaled by $\sin\theta$ as a function of bias, respectively. The in-plane torkance is almost a constant within a small bias, which is consistent with theoretical predictions\cite{Slonczewski_JMMM_1996}. The in-plane torkances under different noncollinear angles almost overlap with each other, showing a good description of $\sin\theta$ dependence, as do the out-of-plane ones. Such an angular dependence of STT is robust. As shown theoretically in Ref.~\onlinecite{Wilczynski_PRB_2008}, changing the lead polarization or the width and height of the insulating layer does not change qualitatively the angular dependence in a MTJ system. Actually, the $\sin \theta$ angular dependence is common in magnetic systems regardless of whether there is a metallic spacer\cite{Slonczewski_JMMM_1996,Alexey_PRB_2007,Brataas_NatMat_2012} or an insulating spacer\cite{Wilczynski_PRB_2008}. However, it is worth noting that Yu $et~al.$ showed that angular dependence would deviate from the standard $\sin \theta$ form under large bias.\cite{YuYunjin_Nanotech2013}

    Now we investigate ac modulation effects in this CNT-based MTJ device. One may wonder how much would the STTs be changed and whether the dc bias dependence and angular dependence would be altered or not. To avoid adding up the total bias, we suppose that both leads have the same ac modulation amplitude, i.e., $\Delta_L=\Delta_\textrm{R}\equiv \Delta$. In this case, no ac bias is applied and electrons are mainly driven by the dc bias under ac modulation.\footnote{Setting $\Delta_L=-\Delta_R$ leads to the same time-averaged results. However, there is instant bias increase, which we try to avoid with.}
    Firstly, we study the STTs under ac modulation of $\Delta=\omega$, and we demonstrate the results in Figs.~\ref{fig3}(a) and (b). From the figures, one can see that as $\Delta$ and $\omega$ increase, both STTs increase obviously in the magnitude. Also, the linear bias and quadratic bias dependence within the small-bias region of the in-plane and out-of-plane STTs seem to maintain well, respectively, together with the $\sin\theta$ angular dependence (inset figures).

    Given the fact that spin currents here are carried by electrons, enhancing transmission of charge carriers will naturally be accompanied by the enhancement of spin currents.
    In this semiconducting device that we study, it is necessary to get contributions from electrons away from the Fermi energy to enhance both charge and spin transport. As mentioned above, the nearest transmission peaks are located at $\pm \varepsilon_0$ with $\varepsilon_0\approx 0.44$~eV. By ac modulation, it is possible to get these peaks involved in transport, resulting in an enhancement.

    Following this thought, we compute STTs as a function of $\Delta$ using a driving ac frequency $\omega=\varepsilon_0/k$ with $k=1,2,3$ under a small bias (up to 0.1 V) to get the linear (quadratic) coefficients $\alpha$($\beta$) of the in-plane (out-of-plane) STT. Larger coefficients means that larger STTs are obtained under a given dc bias. As shown in Fig.~\ref{fig4}, the results at $\Delta=0$ correspond to the dc limit. As shown in Fig.~\ref{fig4}(a), $\alpha$ is generally much bigger than that under a dc bias. For the case of $\Delta/\omega=6$ and $\omega=\varepsilon_0$, the improvement is about 12 times. 
    Compared to the in-plane component, $\beta$ of the out-of-plane component changes significantly with $\Delta$--not only does it change its amplitude by up to 75 times (P$_1$ in Fig.~\ref{fig4}(b)), but it also may change its sign. This means that the bias-induced out-of-plane torque may have a significant enhancement and even sign reversal under a resonant driving frequency. The anomalous dependence on $\Delta$ should be able to be detected by ferromagnetic resonance experiments.\cite{Slonczewski_Nat_2007,zhaohui}
    For the case of $\omega=\varepsilon_0/3$, the first peak of the out-of-plane STT occurs around $\Delta/\omega=4.2$, where is the exact location of the first peak of the Bessel function $J_3(x)$, denoted as P$_3$ in the figure. From Eq.~(\ref{eq:Js_xzSim}), this means that there is a considerable contribution coming from $E-3\omega$. Thus, we have a strong evidence that this ac enhancement is mainly contributed by 3-photon-assisted tunneling. Similarly, the first peaks of the out-of-plane STT as a function of $\Delta/\omega$ when $\omega=\varepsilon_0/2$ and $\omega=\varepsilon_0$ coincide with P$_2$ and P$_1$, implying 2-photon-assisted and 1-photon-assisted tunneling, respectively. From the above results, it implies that using an ac driving frequency $\omega=\varepsilon_0/k,\, k=1,2,\cdots$ together with $\Delta/\omega=P_k, k=1,2,\cdots$ stimulates the best performances of photon-assisted tunneling STTs by involving remote transmission channels. Further analysis reveals that the in-plane STT tends to increase more slowly than the charge current. Meanwhile, the out-of-plane torque may increase faster or slower than charge current, depending on AC modulation parameters. Therefore, although it is capable of increasing spin transfer torque, AC modulation does not increase the amount of spin torque delivered per electron.

    As indicated by Eqs.~(\ref{eq:Js_xzSim}-\ref{eq:Js_ySim}), there are two factors influencing the ac modulation output: weighting coefficients $J_k^2(\Delta/\omega)$ and spin transmission coefficients at $E-k\omega$. When $\omega$ is fixed as previously, variation of $\Delta$ only affects the weighting coefficients and thus the enhancement in STTs can be simply optimized by finding out the peak values of $J_k^2(\Delta/\omega)$. However, when $\omega$ changes, both coefficients change with $\omega$ and the variation is complicated. Figures~\ref{fig5}(a) and (b) illustrate STTs as functions of $\Delta$ and $\omega$ ($\Delta=0$ corresponds to dc cases). This shows that ac modulation generally has a large impact on both STTs as $\Delta$ grows. Contrary to Ref.~\cite{SuGang_PRB2003}, where the variation of $\omega$ seems to have no impact on the in-plane STT, STTs here show a strong dependence on the value of $\omega$, which is attributed to the involvement of transport channels from the sandwiched material of the MTJ. Introducing the enhancement factor $\eta _{||,\perp}$ as
    \begin{align}
    {\eta _{||, \bot }}({V_b}, \Delta, \omega) = \frac{{{\tau _{||, \bot }}({V_b},\Delta ,\omega ) - {\tau _{||, \bot }}(0,\Delta ,\omega )}}{{\tau _{||, \bot }^{dc}({V_b}) - \tau _{||, \bot }^{dc}(0)}},
    \end{align}
    we plot $\eta _{||,\perp}$ at $V_b=0.01$ V and $\Delta=0.619$ eV as a function of ac driving frequency $\omega$ in Fig.~\ref{fig5}(c).
    In this situation, the ac enhancement factor of the in-plane STT is about 3$\sim$4, while that for the out-of-plane STT ranges from -75 to 30.

    One may expect the best enhancement to occur at $\omega=\varepsilon_0/k,~k=1,2,\cdots$ contributed by enhanced spin transmission, as discussed previously. However, optimized enhancements are not necessarily located at these special points. Decomposed contribution analysis in Fig.~\ref{fig5}(d) offers further details about the main contribution. For example, the in-plane STT at $\omega=0.1006$ eV, which satisfies $\varepsilon_0/5 < \omega < \varepsilon_0/4$ , is mainly contributed by 4- and 5-photon-assisted tunneling. This shows that peaks of the in-plane STT between $\omega=\varepsilon_0/k$ and $\varepsilon_0/(k+1)$ are highly likely to be contributed by $k$- and ($k$+1)-photon-assisted tunneling, indicating a compromise between the simultaneous variation of $J_k^2$ and spin transmission. For the out-of-plane STT, however, the best enhancement occurs around $\omega=\varepsilon_0/k,~k\in \mathbb{N}$ [black dots in Fig.~\ref{fig5}(c)], which implies that changes in spin transmission dominates. Indeed, decomposition of individual contributions in Fig.~\ref{fig5}(d) distinctly supports our argument. With this decomposition, it is shown that resonant tunneling assisted by the $k$-photon process significantly changes STTs.

\section{Conclusion \label{sec:conclusion}}
    In conclusion, by deriving the formulation of time-averaged spin transfer torque in MTJs under ac harmonic modulation using the nonequilibrium Green's function method within the wide-band-limit approximation, we are able to investigate ac modulation of STTs in MTJs. Using a (5,5) CNT as an example, we show that STTs under ac modulation maintain the basic features of low-bias linear (in-plane STT) and quadratic (out-of-plane STT) dependence, and the $\sin\theta$ angular dependence. And remarkably, by tuning the ac driving frequency to be at resonant frequencies of $k$-photon-assisted tunneling and tuning the ac amplitude $\Delta$ to maximize the weighting factor $J_k^2(\Delta/\omega)$, we are able to enhance the bias-induced in-plane (out-of-plane) STTs by up to about 12 times (75 times).

    In the above, we focus on a symmetric system in which harmonic modulation on both leads is the same. However, Eq.~(\ref{eq:aveJs}) is also applicable to cases in which $\Delta_L \ne \Delta_R$. In those cases, the quadratic dependence of the out-of-plane torque on bias can be changed. Also, there are still rich physics underneath, such as asymmetric junctions, leads with different polarization, carbon nanotubes of different chirality and length, and also defects. We expect our method to be applicable in MTJ systems with ferromagnetic leads, and our proposal of ac modulation of STTs offers a way to reduce operating bias of STT devices, showing promising applications in future nanoelectronics and spintronics.

\begin{acknowledgments}
We gratefully acknowledge financial support by the Natural Sciences and Engineering Research Council of Canada (NSERC) of Canada (H.G.). We thank Calcul Qu\'ebec and Compute Canada for the computation facilities.
\end{acknowledgments}

\appendix
\renewcommand\thefigure{\thesection.\arabic{figure}}

\section{Spin current}\label{app1}
Let us consider a general Hamiltonian
    \begin{align}
    H = \sum\limits_{i,s=\pm } {\left( {{\varepsilon _{is }}\hat{ c}_{is }^\dag {\hat{ c}_{is }} + {M_i}\hat{ c}_{is }^\dag {\hat{ c}_{i\bar s }}} \right)}  + \sum\limits_{ < i,j > ,s s'} {{t_{is ,js '}}\hat{ c}_{is }^\dag {\hat{ c}_{js '}}} ,
    \end{align}
    where $s=+,-$ labels the spin eigenstates along the $z$ direction, and ${\varepsilon _{is }}$ may depend on time, spin index, and site. According to the definition of spin current (Eq.~(\ref{eq:def}) in the main text), \textcolor{blue}{the net spin current flux into site $i$} is

    \begin{widetext}
    \begin{align}
      \left( \frac{{d{{\bm{\hat{ s}}}_i}}}{{dt}}\right)_{hopping}
    & = \frac{1}{i\hbar} \frac{\hbar}{2} \sum_{s's''} {\bm{\sigma} _{s's''}}\left[ {c_{is'}^\dag {c_{is''}},\sum\limits_{mk,ss'''} {{t_{ms,ks'''}}c_{ms}^\dag {c_{ks'''}}} } \right]  \cr
    & = \frac{1}{{2i}}\sum\limits_{<j,i>, s s' s''} {} {\bm{\sigma} _{s' ,s''}}  \left( {{t_{is'',js}} {\hat{c}_{is' }^\dag {\hat{c}_{js}}}   - {t_{js,is' }} {\hat{c}_{js}^\dag {c_{is''}}}  } \right).
    \end{align}
    \end{widetext}
    Defining Green's function as
    \begin{align}
     G_{is' ,js}^ < \left( {t,t'} \right) = i\left\langle {\hat{ c}_{js}^\dag \left( {t'} \right){\hat{ c}_{is' }}\left( t \right)} \right\rangle /\hbar  ,
    \end{align}
we have
    \begin{align}
    {\left[ {G_{is' ,js}^ < \left( {t,t} \right)} \right]^\dag } =  - i\left\langle {\hat{ c}_{is' }^\dag \left( t \right){\hat{ c}_{js}}\left( t \right)} \right\rangle /\hbar =  - G_{js,is' }^ < \left( {t,t} \right),
    \end{align}
and the above spin current turns out to be
    \begin{widetext}
    \begin{align}\label{eq:Js1}
    \left( \frac{{d{{\bm{\hat{ s}}}_i}}}{{dt}}\right)_{hopping}  
     &=   \frac{1}{{2i}}\sum\limits_{<j,i>,s s' s''} {} {\bm{\sigma} _{s' ,s''}}  \left( {{t_{is'',js}}\left\langle {\hat{c}_{is' }^\dag {\hat{c}_{js}}} \right\rangle  - {t_{js,is' }}\left\langle {\hat{c}_{js}^\dag {c_{is''}}} \right\rangle } \right) \cr
     &= -\frac{\hbar }{2}\sum\limits_{<j,i>,s s' s''} {} {\bm{\sigma} _{s' ,s''}} {\left[ {{t_{is'',js}}G_{js,is' }^ < \left( {t,t} \right) - {t_{js,is' }}G_{is'',js}^ < \left( {t,t} \right)} \right]} \cr
     &= -\frac{\hbar }{2}\sum\limits_{<j,i>,s s' s''}^{} {\left[ {{\bm{\sigma} _{s' ,s''}}{t_{is'',js}}G_{js,is' }^ < \left( {t,t} \right) - {t_{js,is' }}{\bm{\sigma} _{s' ,s''}}G_{is'',js}^ < \left( {t,t} \right)} \right]}
    \end{align}
    \end{widetext}

    The spin current flowing into lead R can be obtained by considering the total spin change in the lead R caused by hopping terms as:
    \begin{align}\label{eq:Js_C2R}
        {\bf{J}}_{\textrm{C} \to \textrm{R}}^s\left( t \right)
         =& \sum\limits_{i \in R}^{} {} {\left\langle {\frac{{d{{\bf{s}}_i}}}{{dt}}} \right\rangle _{hopping}}\cr
         = & - \frac{\hbar }{2}\sum\limits_{\scriptstyle s s' s'', < i,j > , i
         \in \textrm{R},j \in \textrm{C}\hfill}^{} {\left[ {{{\bm{\sigma }}_{s' ,s''}}{t_{is'',js}}G_{js,is' }^ < \left( {t,t} \right)} \right.} \cr
         & {\left. {  - {t_{js,is' }}{{\bm{\sigma }}_{s' ,s''}}G_{is'',js}^ < \left( {t,t} \right)} \right]}
    \end{align}

    When lead R has ${\bf{M}}_\textrm{R}$ along $z$, and hopping between the central region and the lead R does not cause spin-flipping, we have
    \begin{widetext}
    \begin{align}
    {\bf{J}}_{\textrm{C} \to \textrm{R}}^s\left( t \right)
    &=  - \frac{\hbar }{2}\sum\limits_{s s', < i,j > ,i \in \textrm{R},j \in \textrm{C}}^{} {\left[ {{{\bm{\sigma }}_{s' ,s}}{t_{i,j}}G_{js ,is' }^ < \left( {t,t} \right) - {t_{j,i}}{{\bm{\sigma }}_{s' ,s}}G_{is,js' }^ < \left( {t,t} \right)} \right]} \cr
    & =  - \frac{\hbar }{2}\sum\limits_{s s ', < i,j > ,i \in \textrm{R},j \in \textrm{C}}^{} {\left[ {{{\bm{\sigma }}_{s' ,s}}{t_{i,j}}G_{js,is' }^ < \left( {t,t} \right) + t_{i,j}^*{{\bm{\sigma }}_{s,s' }}G_{js,is' }^{ < *}\left( {t,t} \right)} \right]} \cr
    & =  - \frac{\hbar }{2}\sum\limits_{s s ', < i,j > ,i \in \textrm{R},j \in \textrm{C}}^{} {\left[ {{{\bm{\sigma }}_{s' ,s}}{t_{i,j}}G_{js,is' }^ < \left( {t,t} \right) + t_{i,j}^*{\bm{\sigma }}_{s' ,s}^*G_{js,is' }^{ < *}\left( {t,t} \right)} \right]} \cr
    & =  - \hbar \sum\limits_{s s ', < i,j > ,i \in \textrm{R},j \in \textrm{C}}^{} {{\mathop{\rm Re}\nolimits} \left[ {{{\bm{\sigma}}_{s' ,s}}{t_{i,j}}G_{js ,is' }^ < \left( {t,t} \right)} \right]} \cr
     &=- \hbar\sum\limits_{\hfill \scriptstyle s s ' \hfill\atop \scriptstyle {\bf{k}}\alpha  \in \textrm{R}, n\in \textrm{C} \hfill}{} {{\rm{Re}}\left[ {G_{ns ,{\bf{k}}\alpha s'}^ < \left( {t,t} \right){t_{{\bf{k}}\alpha ,n}}{{\bm{\sigma }}_{s',s }}} \right] }.
    \end{align}
    \end{widetext}

\section{Simplification of the formulation under the chosen coordination} \label{app2}
   Under the coordination shown in Fig. 1 in the main text, $\Gamma_L$ and $\Gamma_R$ are always real (see Eqs.~(\ref{eq:GammaR}) and (\ref{eq:GammaL}) in the main text). Also, both Hamiltonians of the central region and the hopping integrals between leads and the central region are real. Given these conditions, it can be proven that
   \begin{align}
        {\rm{Im}}\left( {G_k^r} \right)
        &=  - \frac{1}{2}G_k^r\Gamma G_k^a, \\
        {\rm{Im}}G_k^ <
        &= \sum\limits_\alpha ^{} {{f_\alpha }} G_k^r{\Gamma _\alpha }G_k^a.
   \end{align}
   Therefore, we have
   \begin{align}
        &{\rm{Im}}\left( {G_k^r{\Gamma _\textrm{R}}{{\rm{\sigma }}_{x/z}}} \right) \cr
        &= {\rm{Im}}\left( {G_k^r} \right){\Gamma _\textrm{R}}{\mathop{\rm Re}\nolimits} \left( {{{\rm{\sigma }}_{x/z}}} \right) + {\rm{Re}}\left( {G_k^r} \right){\Gamma _\textrm{R}}{\rm Im} \left( {{{\rm{\sigma }}_{x/z}}} \right)\cr
        &= {\rm{Im}}\left( {G_k^r} \right){\Gamma _\textrm{R}}{{\rm{\sigma }}_{x/z}}\cr
        & =  - \frac{1}{2}G_k^r\left( {{\Gamma _\textrm{L}} + {\Gamma _\textrm{R}}} \right){\rm{G}}_k^a{\Gamma _\textrm{R}}{{\rm{\sigma }}_{x/z}},
    \end{align}
    \begin{widetext}
    \begin{align}
        & J_{k;\textrm{R}}^2{\rm{ImTr}}\left( {G_k^r{\Gamma _\textrm{R}}{{\bf{\sigma }}_{\bf{\nu }}}} \right){f_\textrm{R}} +
        \frac{1}{2}\sum\limits_{\alpha  = \textrm{L,R}} {J_{k;\alpha }^2} {f_\alpha }{\rm{ReTr}}\left( {G_k^r{\Gamma _\alpha }G_k^a{\Gamma _\textrm{R}}{\sigma _{x/z}}} \right)\cr
        & =  - \frac{1}{2}J_{k;\textrm{R}}^2{f_\textrm{R}}{\rm{Tr}}\left[ {G_k^r\left( {{\Gamma _\textrm{L}} + {\Gamma _\textrm{R}}} \right){\rm{G}}_k^a{\Gamma _\textrm{R}}{\sigma _{x/z}}} \right]
        + \frac{1}{2}\sum\limits_{\alpha  =\textrm{ L,R}} {J_{k;\alpha }^2} {f_\alpha }{\rm{ReTr}}\left( {G_k^r{\Gamma _\alpha }G_k^a{\Gamma _\textrm{R}}{\sigma _{x/z}}} \right)\cr
        & = \frac{1}{2}\left( {{f_\textrm{L}}J_{k;\textrm{L}}^2 - {f_\textrm{R}}J_{k;\textrm{R}}^2} \right){\rm{ReTr}}\left( {G_k^r{\Gamma _\textrm{L}}G_k^a{\Gamma _\textrm{R}}{\sigma _{x/z}}} \right)
    \end{align}
    \end{widetext}
    and then
        \begin{align}
        J^s_{x/z}
        &=\int_{}^{} {\frac{{d\varepsilon }}{{2\pi }}} \sum\limits_k {J_{k;\textrm{R}}^2} {\rm{ImTr}}\left( {G_k^r{\Gamma _\textrm{R}}{{\bf{\sigma }}_{\bf{\nu }}}} \right){f_\textrm{R}}\left( \varepsilon  \right) + \cr
        & \sum\limits_{\alpha  = \textrm{L,R}} {} \int {\frac{{d\varepsilon }}{{4\pi }}} {f_\alpha }\left( \varepsilon  \right)\sum\limits_k {J_{k;\alpha }^2} {\rm{ReTr}}\left( {G_k^r{\Gamma _\alpha }G_k^a{\Gamma _\textrm{R}}{\sigma _\nu }} \right)\cr
        &= \int_{}^{} {\frac{{d\varepsilon }}{{4\pi }}} \sum\limits_k {} \left( {{f_\textrm{L}}J_{k;\textrm{L}}^2 - {f_\textrm{R}}J_{k;\textrm{R}}^2} \right){\rm{ReTr}}\left( {G_k^r{\Gamma _L}G_k^a{\Gamma _R}{\sigma _{x/z}}} \right)\cr
        \end{align}
        When ${{\Delta _\textrm{L}} = {\Delta _\textrm{R}}}$, it turns to be
        \begin{align}
         J_{x/z}^s = \int_{}^{} {\frac{{d\varepsilon }}{{4\pi }}} \sum\limits_k {} J_k^2\left( {{f_\textrm{L}} - {f_\textrm{R}}} \right){\rm{ReTr}}\left( {G_k^r{\Gamma _\textrm{L}}G_k^a{\Gamma _\textrm{R}}{\sigma _{x/z}}} \right).
        \end{align}

        For $y$ component when ${\Delta _\textrm{L}} = {\Delta _\textrm{R}}$,
        we have
        \begin{align}
        &\sum\limits_{\alpha  \in \textrm{ L,R}} {} \int {\frac{{d\varepsilon }}{{4\pi }}} {f_\alpha }\left( \varepsilon  \right)\sum\limits_k {J_{k;\alpha }^2} {\rm{ReTr}}\left( {G_k^r{\Gamma _\alpha }G_k^a{\Gamma _R}{\sigma _y}} \right)\cr
        & =\int {\frac{{d\varepsilon }}{{4\pi }}} \sum\limits_k {J_k^2} {\rm{Tr}}\left\{ {{\rm{Re}}\left[ {\sum\limits_{\alpha  \in \textrm{ L,R}} {{f_\alpha }\left( \varepsilon  \right)} G_k^r{\Gamma _\alpha }G_k^a} \right] } \right. \cr
        &\left. {{\mathop{\rm Re}\nolimits} \left( {{\Gamma _\textrm{R}}} \right){\mathop{\rm Re}\nolimits} \left( {{\sigma _y}} \right)} \right\}\cr
        & = 0
        \end{align}
        so that
        \begin{align}
         J_{y}^{s}
        &= \int_{}^{} {\frac{{d\varepsilon }}{{2\pi }}} \sum\limits_k {J_{k;\textrm{R}}^2} {\rm{ImTr}}\left( {G_k^r{\Gamma _\textrm{R}}{{\bf{\sigma }}_y}} \right){f_\textrm{R}}\left( \varepsilon  \right)\cr
        &+ \frac{1}{2}\sum\limits_{\alpha  =\textrm{ L,R}} {} \int {\frac{{d\varepsilon }}{{2\pi }}} {f_\alpha }\left( \varepsilon  \right)\sum\limits_k {J_{k;\alpha }^2} {\rm{ReTr}}\left( {G_k^r{\Gamma _\alpha }G_k^a{\Gamma _\textrm{R}}{\sigma _y}} \right)\cr
        & = \int_{}^{} {\frac{{d\varepsilon }}{{2\pi }}} \sum\limits_k {J_{k;\textrm{R}}^2} {\rm{ImTr}}\left( {G_k^r{\Gamma _\textrm{R}}{{\bf{\sigma }}_y}} \right){f_\textrm{R}}\left( \varepsilon  \right)
        \end{align}

Meanwhile, charge current can be obtained by omitting $\sigma_z$ in spin current $ J_z^s$, and multiplying a prefactor $e/(\hbar/2)$ as:
    \begin{align}
    {{ J}_c} &= \frac{e}{{\hbar /2}}\cdot \frac{1}{2}\int_{}^{} {\frac{{d\varepsilon }}{{2\pi }}} \sum\limits_k {\left( {{f_\textrm{L}}J_{k;\textrm{L}}^2 - {f_R}J_{k;\textrm{R}}^2} \right)} {\rm{Tr}}\left( {G_k^r{\Gamma _L}G_k^a{\Gamma _\textrm{R}}} \right)\cr
    & = \frac{e}{h}\int_{}^{} {d\varepsilon } \sum\limits_k {\left( {{f_\textrm{L}}J_{k;\textrm{L}}^2 - {f_\textrm{R}}J_{k;\textrm{R}}^2} \right)} {\rm{Tr}}\left( {G_k^r{\Gamma _\textrm{L}}G_k^a{\Gamma _\textrm{R}}} \right)
    \end{align}

%

\end{document}